\documentstyle[amssymb,twocolumn,aps,epsfig]{revtex}

\begin{document}
\twocolumn[\hsize\textwidth\columnwidth\hsize\csname @twocolumnfalse\endcsname
\draft
\title{c-axis Josephson Tunneling in Twinned YBCO Crystals}
\author{Robert Haslinger$\dagger$ and Robert Joynt$\dagger \ddagger$}
\address{$\dagger$ Department of Physics, University of Wisconsin Madison, \\
  1150 University Avenue, 
  Madison, WI 53706, USA}
\address{$\ddagger$ Center for Theoretical Sciences, P.O. Box 2-131, Hsinchu,
Taiwan 300, Republic of China}
\date{\today}
\maketitle

\begin{abstract}
Josephson tunneling between 
YBa$_2$Cu$_3$O$_{7-\delta}$ and Pb with the current
flowing along the c-axis of the YBa$_2$Cu$_3$O$_{7-\delta}$
is presumed to come from an s-wave component of the 
superconductivity in YBa$_2$Cu$_3$O$_{7-\delta}$.  Experiments
on multi-twin samples are not entirely consistent
with this hypothesis.  The sign changes of the s-wave order parameter
across the $N_T$ twin boundaries should give cancellations,
resulting in a small ($\sqrt{N_T}$) tunneling current.  
The actual current is larger than this.  We present a theory 
of this unexpectedly large current based upon
a surface effect: disorder-induced supression of the d-wave 
component at the (001) surface leads to s-wave coherence
across the twin boundaries and a non-random
tunneling current.  We solve the case of an ordered array of
d+s and d-s twins, and estimate that the twin size at which  
s-wave surface coherence occurs is consistent with typical
sizes observed in experiments.  In this picture, there is
a phase difference of $\pi/2$ between different surfaces
of the material.  We propose a corner junction experiment
to test this picture.
\end{abstract}
\vspace*{\baselineskip}

]

\section{\protect\bigskip Introduction}
Understanding the nature of the order parameter is one of the main
challenges in the theory of high-T$_{c}$ superconductivity. \ One of the
most fundamental issues is the competition between s-wave and d-wave. \
d-wave is definitely the rule for high-T$_{c}$ systems, yet there is strong
evidence that the electron-doped NdCeCuO material is s-wave. \ At a
microscopic level, this suggests that the mean-field pairing interaction has
two eigenvalues which vie for dominance. \ Understanding this competition
would provide insight into the pairing mechanism.

In this regard YBa$_{2}$Cu$_{3}$O$_{7-\delta }$ (YBCO) is of particular
interest. \ In other systems, the square symmetry forces the ordering to be
pure d-wave or pure s-wave. \ The presence of one most likely prevents the
emergence of the other because of repulsive terms in the free energy, and
the competition has a clear winner. \ \ In contrast, YBCO has orthorhombic
symmetry. \ This makes it inevitable that d-wave and s-wave are always mixed 
\cite{lkj}. \ Josephson tunneling experiments with current flowing mainly in
the a-b plane \cite{vh} have made it clear that the dominant component is
d-wave, but a substantial body of work has also demonstrated the existence
of Josephson tunneling along the c-axis from YBCO to a Pb electrode,
indicative of an s-wave component \cite{sun} \cite{sun2}\cite{kouz}\cite
{Kleiner}. \ It is to be hoped that these latter experiments, if carefully
analyzed, can tell us about the strength of the s-wave admixture in the
order parameter.

We shall first review the experimental and theoretical situation concerning
c-axis tunneling, concluding that there are major theoretical puzzles still
to be resolved. \ Then we shall present a new model of the phenomena which
we argue is in agreement with the data as they stand, and how to test the
model more thoroughly. \ 

c-axis tunneling from YBCO to Pb was first observed in twinned crystals
etched wth Br \ \cite{sun}, with an $I_{c}R_{n}$ product of as much as 10 \%
of the known gap of about 30 meV. \ This strongly suggested the presence of
an s-wave component of the superconductivity of YBCO, as a pure d-wave
current would average to zero over the Fermi surface. However, another
possibility was that the current was due to second order tunneling of the
d-wave component \cite{jpn}. \ This hypothesis predicts the presence of
Shapiro steps in the conductivity in units of $hf/4e$, where $f$ is the
frequency of the incident radiation. \ This was ruled out in subsequent
microwave experiments \cite{Kleiner} \ \ Finally, the question of tunneling
through step walls at the surface arises, particularly if it is deeply
etched. \ This would be a process in which the current actually flows in the 
$a-b$ direction. \ However, the fabrication of $a-b$ junctions \cite{sun2}
and the observation of tunneling {\it in situ} without etching \cite{lesueur}
appears to have laid this possibility to rest. \ The presence of a nonzero
s-wave component in YBCO must now be accepted.

Is it reasonable to accept the 10\% estimate of $s$ to $d$ which comes from
the $I_{c}R_{n}$ product at face value? \ Clearly not, for the following
reason. \ A twinned sample should have a relative population of twins of the
two possible orientations given by statistical considerations. \ The d-wave
component must remain coherent through the sample, as is shown by the corner
junction experiments \cite{vh}. \ Because the change in orientation reverses
the relative sign of $s$ and $d$ we should have roughly equal areas of $d+s$
and $d-s$ superconductivity in the sample, in which case the net current
should be zero. \ \ More precisely, the net current should be proportional
to to $I_{c}\sqrt{N_{T}}$, where $I_{c}$ is the critical current of a single
twin and $N_{T}$ is the total number of twins. \ If we accept this argument,
then the actual proportion of $s$ to $d$ would be higher than 10\%. \ This
would move the nodes in the gap well away from the diagonal in the Brillouin
zone. \ This would be inconsistent with tricrystal experiments \cite{tri} \
Furthermore, comparison of Josephson currents in single crystals to twinned
crystals show similar $R_{n}$values and $I_{c}$ values which range from 0.5
to 1.6 mA for single crystals and from 0.1 to 0.9 mA for twinned samples 
\cite{sun2}. \ These numbers are subject to the objection that one cannot be
sure that the tunneling matrix elements are not extremely sensitive to the
sample preparation method. \ Nevertheless, in view of the fact that $R_{n}$
does not vary wildly from sample to sample, they suggest that a purely
statistical analysis of twin populations with a resulting small imbalance of 
$d+s$ and $d-s$ is not a viable explanation of the data.

The dilemma was deepened by experiments on crystals with much larger twins,
large enough so that junctions could be formed which straddled either one or
even zero twin boundaries \cite{kouz}. \ These showed that the direction of
current definitely did change sign across the twin boundary, a fact which
can be established unambiguously by investigating the current as a function
of field in the plane of the junction. \ This observation was consistent in
all eight samples studied. \ Also, no such sign changes were observed in the
absence of a boundary. \ These experiments therefore clearly confirm the
mechanism of an s-wave component controlled by the orthorhombic distortion,
without offering any explanation of the large current in heavily twinned
samples. \ One further observation in these experiments may offer a clue,
however. \ The current at zero appplied field in single-boundary samples was
consistently higher than calculated by looking at the relative sizes of the
two twins. \ We will return to this point below.

Summarizing the experiments, we may say that an s-wave component which
changes sign across boundaries is clearly present. \ If it always changes
sign, then we cannot explain the data on twinned samples using purely
statistical arguments. \ One possibility is that the twin populations are
not equally likely. \ For example, if the twinning takes place under
uniaxial stress, then one orientation would be favored. \ Experiments which
correlate microstructure with Josephson current are needed to rule this out 
\cite{sun2}. \ However, given the size of the Josephson effect in twinned
samples, it seems to us that this explanation is somewhat implausible. \ 

The most detailed theory of c-axis tunneling proposed to date is that
offered by Sigrist {\it et al.} \cite{sigrist}. Their picture involves no
net tunneling from the twins themselves. A time-reversal-breaking state at
the twin boundary is predicted which results in a net Josephson current
coming from the twin boundaries. This would give a Josephson current which
is proportional to the number of boundaries for a fixed surface area. \ This
is not observed, though again one must keep in mind that different samples
must be compared to make any such statement, and variations in important
microscopic parameters canot be controlled in such comparisons. \
In
addition, however, the theory predicts a current which has maximum asymmetry
(as a function of in-plane angle) when the applied magnetic field is
parallel to the boundary. \ This is an experiment in which the unknown matrix
elements are held fixed. \ This prediction is in conflict with the
experimental observations, which are symmetric at parallel orientation \cite
{Kleiner}. \ 

A quite different proposal was made by Xu {\it et al. }\cite{xu}. \ These
authors postulate a bulk $d+is$ \ state. \ In this theory, however, the
s-component does not change sign across the boundary, which does not agree
with the measurements on single-boundary samples in a parallel field.

We present an alternative explanation in which the nonzero tunneling current
is the result of a surface effect. \ YBCO\ is notorious in photoemission
experiments for not showing a gap. This proposal is inspired by the fact
that photoemission experiments (with resolutions of order $10meV$ or less)
have also never succeeded in seeing a gap at the (001) surface in this
material (in contrast to Bi$_{2}$Sr$_{2}$CaCu$_{2}$O$_{8+x}$). This shows
that the magnitude of the gap at this particular surface is much reduced. \
Furthermore, if this reduction is due to disorder, such as surface
scattering, one would expect that the d-wave component is relatively much
more suppressed than any s-wave admixture. A similar suppression could
result from an oxygen vacancy concentration gradient. \ This suppression of
the d-wave component of the order parameter as we approach the (001) surface
of the YBCO in the context of c-axis tunneling is of two central hypotheses
of our model and was first suggested by Bahcall \cite{bahcall}. The second
crucial ingredient is new: the d-wave surface suppression results in a
coherent s-wave surface layer and hence an enhanced Josephson tunneling
current in highly twinned samples without a very large admixture of s-wave
in the bulk.

\section{Double Twin Model}

\begin{figure}
\begin{center}
\leavevmode
\epsfxsize \columnwidth
\epsffile{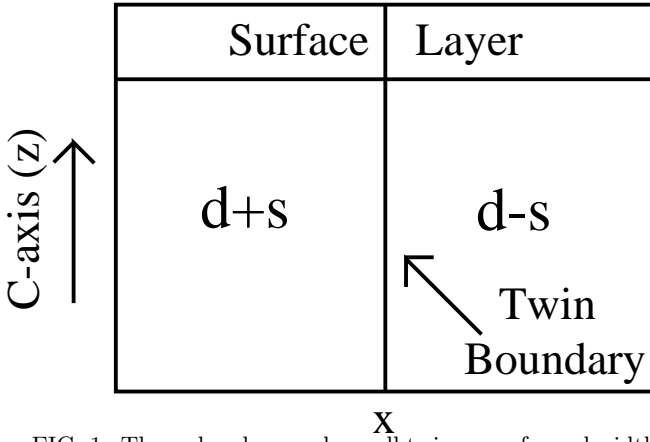}
\caption[]{The ordered case where all twins are of equal width is
equivalent to a d+s and a d-s twin with wrap around boundary conditions.
Near the (001) surface, there is a layer in which the d component of
the order parameter is highly surpressed.}
\label{fig:fig1}
\end{center}
\end{figure}

Twinned samples are disordered on the $\mu m$ scale, the twin boundaries
running predominantly along the diagonal of the a-b plane. It is reasonable
then to approximate the disordered sample by an array of straight twin
boundaries running across the entire sample, which is considered to be
semi-infinite. We concern ourselves in this paper with the ordered case in
which all twins are of the same width and alternate between d+s and d-s. In
real samples the twins have varying widths, but we have verified numerically
that the basic results are unaffected by neglecting the disorder in the
widths. The solution of the ordered model should be periodic with a period
of two twins. Therefore, we solve the case of two twins with periodic
boundary conditions. The twin boundary occupies the half plane defined by $%
x=0$ and $z\leq 0$. The plane $z=0$ is the (001) surface of the YBCO sample.
The model is illustrated schematically in Fig.\ \ref{fig:fig1}.

We shall write the bulk free energy density in terms of the two order
parameters $\Psi _{d}$ and $\Psi _{s}$. Spatial variation is allowed only
along the x-direction, (normal to the twin boundary), and along the
z-direction (normal to the surface). The free energy density is given by: 
\begin{eqnarray}
f &=&\alpha _{d}(z)|\Psi _{d}|^{2}+\frac{\beta _{d}}{2}|\Psi
_{d}|^{4}+K_{dx}|\partial _{x}\Psi _{d}|^{2}+K_{dz}|\partial _{z}\Psi
_{d}|^{2}  \label{fe} \nonumber \\
&&+\alpha _{s}|\Psi _{s}|^{2}+\frac{\beta _{s}}{2}|\Psi _{s}|^{4}
+K_{sx}|\partial _{x}\Psi
_{s}|^{2}+K_{sz}|\partial _{z}\Psi _{s}|^{2}  \nonumber \\
&&+\alpha _{sd}(x)(\Psi _{s}^{\ast }\Psi _{d}+c.c) 
+\beta_{sd}|\Psi _{s}|^{2}|\Psi _{d}|^{2} 
\end{eqnarray}
Some of the important physical ideas behind our model are displayed by this
equation. $\alpha _{d}$ is a function of position in order to enforce the
condition that the d-wave component is suppressed near the surface. Thus $%
\alpha _{d}\rightarrow \alpha _{d0}<0$, a negative constant, as $%
z\rightarrow -\infty $, deep in the bulk. $\alpha _{d}$ increases toward 0
as $z\rightarrow 0$ at the surface. $\alpha _{sd}$ is the s-d coupling
parameter which is a negative constant deep in the $d+s$ twin and positive
constant deep in the $d-s$ twin. $\alpha _{sd}(x)=-\alpha _{sd}(-x)$. \ 

The $\beta _{sd}$ term is the s-d repulsion mentioned in the introduction. \
We will neglect it in the calculations and have included it here only in
order to stress that a large positive $\beta _{sd}$ suppresses all s-d
mixing in the absence of the bilinear $\alpha_{sd}$ term. \ If this term is
present, as it is here because of the orthorhombic distortion, then the size
of the s admixture is controlled by $\alpha_{sd}/\alpha_{d}$. 

We must also include the free energy of the twin boundaries. Any x axis
variation of $\Psi _{s}$ and $\Psi _{d}$ will take place within a distance
of the order of a coherence length about the twin boundary. Since $\xi
_{ab}\approx 20A$ is very small compared to the average twin width ($0.1$ to 
$10\mu m$) we conclude that the detailed structure of the twin boundary is
not very important. We will assume a very thin boundary and thus take $%
\alpha _{sd}$ to be piecewise constant. This is in direct contrast to the
Sigrist {\it et al.}\ model in which the current comes from the twin
boundaries. In our model the current comes from the twins. We therefore
approximate the free energy of the twin boundary by a Josephson-type
coupling: 
\begin{eqnarray}
f_{TB} &=&-J_{s}|\Psi _{s}^{+}||\Psi _{s}^{-}|\cos (\phi _{s}^{+}-\phi
_{s}^{-})  \nonumber \\
&-&J_{d}|\Psi _{d}^{+}||\Psi _{s}^{-}|\cos (\phi _{d}^{+}-\phi _{d}^{-})
\end{eqnarray}
where $J_{s}$ and $J_{d}>0$. We can also now drop the x axis gradient terms
in the bulk free energy. Any x axis variation in the order parameters takes
place near the twin boundary and has been included in the boundary energy.

The problem has been reduced to 2 one-dimensional twins which are Josephson
coupled. However, only one twin is actually required. As the surface of the
YBCO is approached, the {\it magnitude} of $\Psi _{d}$ and $\Psi _{s}$
should vary in exactly the same way in both the $d+s$ and $d-s$ twins. Only
the phases $\phi _{s}$ and $\phi _{d}$ are different. But while the phases
differ between twins, they are not entirely independent. We set $\phi
_{d}=\phi _{s}=0$ in the bulk of the $d+s$ twin, and $\phi _{d}=0$, $\phi
_{s}=\pi $ in the bulk of the $d-s$ twin. As the (001) surface is
approached, variation in $\phi _{s}^{+}$ and $\phi _{s}^{-}$ should be
symmetric about $\pi /2$, while $\phi _{d}^{+}$ and $\phi _{d}^{-}$ will be
symmetric about 0. This allows the boundary energy to be rewritten entirely
in terms of the phases in the $d+s$ twin. 
\begin{eqnarray}
f_{TB}^{d+s}(z) &=&-J_{d}|\Psi _{d}(z)|^{2}\cos (2\phi _{d}(z))  \nonumber \\
&+&J_{s}|\Psi _{s}(z)|^{2}\cos (2\phi _{s}(z))
\end{eqnarray}
$\phi _{s}$ and $\phi _{d}$ in the $d-s$ twin can be deduced immediately,
and the problem is now entirely one-dimensional.

\section{Solution}

We will solve for the order parameters in the $d+s$ twin. The solution for
the $d-s$ twin follows immediately. Our one-dimensional free energy density
is: 
\begin{eqnarray}
f(z) &=&w\{\alpha _{d}(z)|\Psi _{d}|^{2}+\frac{\beta _{d}}{2}|\Psi
_{d}|^{4}+\alpha _{s}|\Psi _{s}|^{2}+\frac{\beta _{s}}{2}|\Psi _{s}|^{4} 
\nonumber \\
&+&K_{dz}|\partial _{z}\Psi _{d}|^{2}+K_{sz}|\partial _{z}\Psi _{s}|^{2} 
\nonumber \\
&+&\alpha _{sd}(x)(\Psi _{s}^{\ast }\Psi _{d}+c.c)\}  \nonumber \\
&+&-J_{d}|\Psi _{d}|^{2}\cos (2\phi _{d})+J_{s}|\Psi _{s}|^{2}\cos (2\phi
_{s})
\end{eqnarray}
where $w$ is the width of a single twin. Performing the usual minimizations,
we get: 
\begin{eqnarray}
\frac{\delta f}{2\delta |\Psi _{d}|} &=&w\{\alpha _{d}(z)|\Psi _{d}|+\beta
_{d}|\Psi _{d}|^{3}+\alpha _{sd}|\Psi _{s}|\cos (\phi _{d}-\phi _{s}) 
\nonumber \\
&+&K_{dz}(-\frac{1}{2}\partial _{z}^{2}|\Psi _{d}|+|\Psi _{d}|^{2}(\partial
_{z}\phi _{d})^{2})\}  \nonumber \\
&-&J_{d}|\Psi _{d}|\cos (2\phi _{d})=0
\end{eqnarray}
and 
\begin{eqnarray}
\frac{\delta f}{2\delta \phi _{d}} &=&w\{-\alpha _{sd}|\Psi _{d}||\Psi
_{s}|\sin (\phi _{d}-\phi _{s})  \nonumber \\
&-&\frac{1}{2}K_{dz}|\Psi _{d}|^{2}\partial _{z}^{2}\phi _{d}\}+J_{d}|\Psi
_{d}|^{2}\sin (2\phi _{d})=0
\end{eqnarray}
The analogous s-wave equations are: 
\begin{eqnarray}
\frac{\delta f}{2\delta |\Psi _{s}|} &=&w\{\alpha _{s}|\Psi _{s}|+\beta
_{s}|\Psi _{s}|^{3}+\alpha _{sd}|\Psi _{d}|\cos (\phi _{s}-\phi _{d}) 
\nonumber \\
&+&K_{sz}(-\frac{1}{2}\partial _{z}^{2}|\Psi _{s}|+|\Psi _{s}|^{2}(\partial
_{z}\phi _{s})^{2})\}  \nonumber \\
&+&J_{s}|\Psi _{s}|\cos (2\phi _{s})=0
\end{eqnarray}
and 
\begin{eqnarray}
\frac{\delta f}{2\delta \phi _{s}} &=&w\{-\alpha _{sd}|\Psi _{d}||\Psi
_{s}|\sin (\phi _{s}-\phi _{d})  \nonumber \\
&-&\frac{1}{2}K_{sz}|\Psi _{s}|^{2}\partial _{z}^{2}\phi _{s}\}-J_{s}|\Psi
_{s}|^{2}\sin (2\phi _{s})=0
\end{eqnarray}

In general, these equations must be solved numerically, but it is instructive
to first consider the limit $K_{dz}=K_{ds}=0$, which may be obtained
analytically. If we consider Eqns.\ 6 and 8 we see that: 
\begin{eqnarray}
&J_{d}&|\Psi _{d}|^{2}\sin (2\phi _{d})=J_{s}|\Psi _{s}|^{2}\sin (2\phi _{s})
\nonumber \\
&=&w\alpha _{sd}|\Psi _{s}||\Psi _{d}|\sin (\phi _{s}-\phi _{d})
\end{eqnarray}
$\phi _{s}$ and $\phi _{d}$ are between $0$ and $\pi /2$. There are only the
two obvious solutions: $\phi _{d}=\phi _{s}=0$ or $\pi /2$.

The particular solution which minimizes the free energy is dependent upon
the relative strengths of the $s$ and $d$ intertwin Josephson couplings, 
{\it i.e.} , the ratio $R=J_{d}|\Psi _{d}|^{2}/J_{s}|\Psi _{s}|^{2}$. If $%
R>1 $ then $\phi _{s}=\phi _{d}=0$. If $R<1$ $\phi _{s}=\phi _{d}=\pi /2$.
This is in the $d+s$ twin. In the $d-s$ twin if $R>1$ we have $\phi _{d}=0$ $%
\phi _{s}=\pi $. If $R>0$ then $\phi _{d}=-\pi /2$ and $\phi _{s}=\pi /2$.
The magnitudes are obtained from the coupled set of equations: 
\begin{eqnarray}
|\Psi _{s}| &=&-\frac{1}{|\alpha _{sd}|}(\alpha _{d}(z)|\Psi _{d}|+\beta
_{d}|\Psi _{d}|^{3})  \nonumber \\
|\Psi _{d}| &=&-\frac{1}{|\alpha _{sd}|}(\alpha _{s}|\Psi _{s}|+\beta
_{s}|\Psi _{s}|^{3})
\end{eqnarray}
where we have assumed that the intertwin coupling has little effect on the
magnitude of the order parameters, that is $J_{s}<<w|\alpha _{s}|$ where $w$
is the twin width. The exact form of $|\Psi _{s}|$ and $|\Psi _{d}|$ will
depend on the $\alpha _{d}(z)$ chosen.

\begin{figure}
\begin{center}
\leavevmode
\epsfxsize \columnwidth
\epsffile{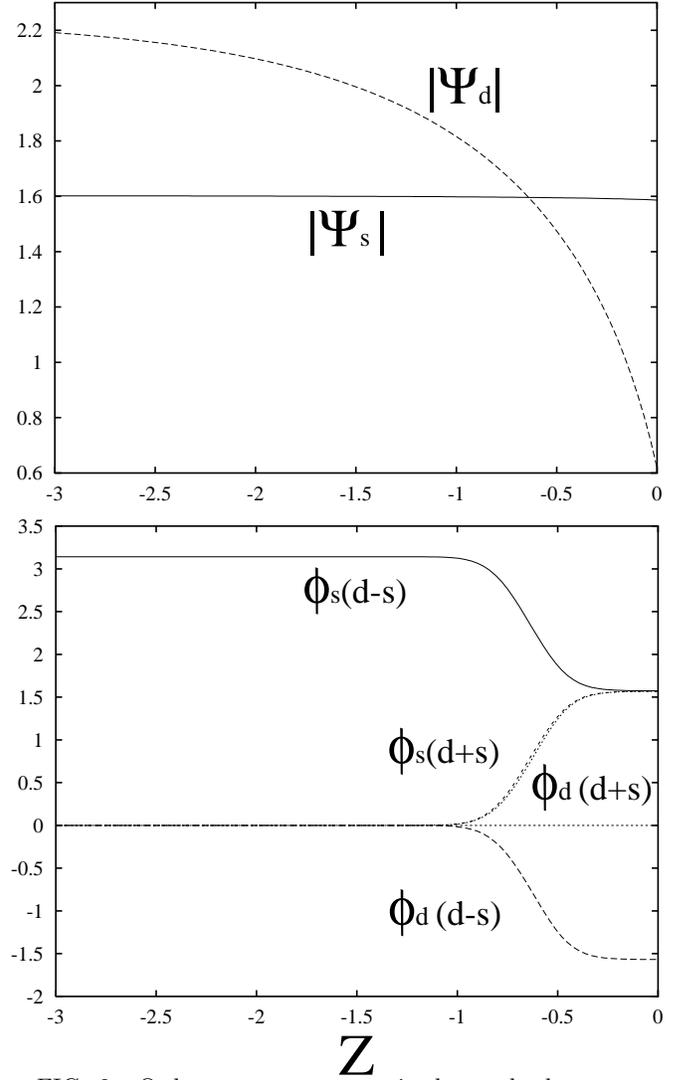}
\caption[]{Order parameter magnitudes and phases near the (001) surface
(z=0). $\alpha_d(z)= -0.95(1-e^z) + 0.05$, $\alpha_s=-0.5$, 
$\beta_d=\beta_s=0.2$, $K_d=K_s=10^{-6}$, $|\alpha_{sd}|=0.01$,
$J_d=J_s=0.005$.  In the bulk $\phi_d$ is coherent across the twin boundary.
At the surface, $\phi_s$ is coherent and is out of phase with the bulk
$\phi_d$ by $\pi/2$.}
\label{fig:fig2}
\end{center}
\end{figure}

The main effect of finite gradient terms $K_{s,d}|\partial _{z}\Psi
_{s,d}|^{2}$ in the free energy is to smooth out the variation in the order
parameters as the surface is approached. We expect the order parameter
magnitudes to be only slightly affected by the introduction of the gradient
terms. Variation in $|\Psi _{d}|$ and $|\Psi _{s}|$ should depend
predominantly on $\alpha _{d}(z)$, since $K_{d}<<|\alpha _{d}|$ etc. The
effect on the phases is more dramatic. For relatively narrow twins, $\phi
_{s}$ and $\phi _{d}$ now undergo a smooth transition from $\phi _{s}=\phi
_{d}=0$ in the bulk of the $d+s$ twin to $\phi _{s}=\phi _{d}=\pi /2$ at the
surface. In the $d-s$ twin, $\phi _{s}$ changes from $\pi $ to $\pi /2$ at
the surface and $\phi _{d}$ from 0 to $-\pi /2$. The order parameter
magnitudes and phases for a model $\alpha _{d}(z)$ are shown in Fig.\ \ref
{fig:fig2}.

The degree of smoothing depends upon the strength of the gradient term
versus that of the coupling across the twin boundary. The dominant factor in
this competition between the gradient and the intertwin coupling energy is
the twin width. For very wide twins, the change in surface phase is
diminished and may be eliminated altogether.

\begin{figure}
\begin{center}
\leavevmode
\epsfxsize \columnwidth
\epsffile{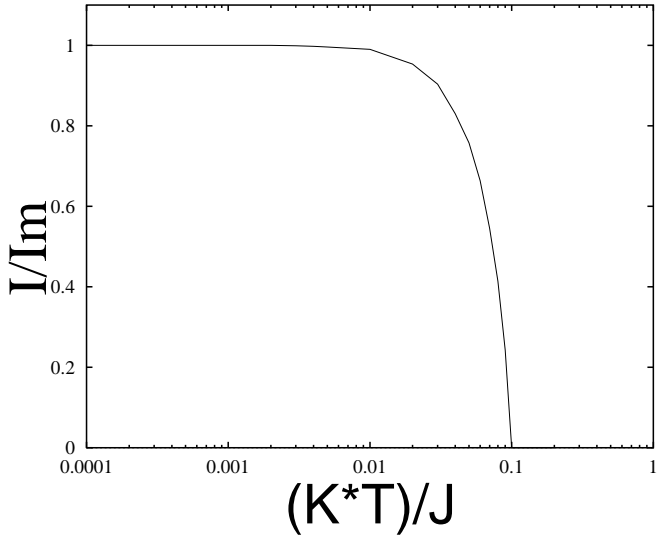}
\caption[]{Net Josephson current vs twin size.  Ginsburg Landau
parameters are as in Fig 2.  $J=J_s=J_d$ and $K=K_s=K_d$. 
T is the twin size.}
\label{fig:fig3}
\end{center}
\end{figure}

The maximum c-axis Josephson current is given by: 
\begin{eqnarray}
\frac{J_{max}}{A} &=&\frac{J_{0}}{2}\{\sin (\phi _{Pb}-\phi _{s}^{d+s})+\sin
(\phi _{Pb}-\phi _{s}^{d-s})\}  \nonumber \\
&=&J_{0}\sin (\phi _{s}^{d+s})
\end{eqnarray}
where A is the junction area and $\phi _{Pb}$ has been chosen to yield the
maximum Josephson current. For very large twins it is not energetically
favorable for the phase change to occur. The s-wave phase at the surface
alternates between $0$ and $\pi $ across twin boundaries and no net
Josephson current flows. As the twins become narrower, a threshold is
reached where the s-wave phases start to shift towards $\pi /2$ at the
surface. The s-wave surface phase alternates between $\phi _{s}^{d+s}$ and $%
\phi _{s}^{d-s}=\pi -\phi _{s}^{d+s}$. Some Josephson coupling is now
possible. For very narrow twins $\phi _{s}$ is coherent across the entire
surface of the crystal, and the maximum Josephson current flows. The current 
{\it saturates}, and further reduction of twin size has no effect on the
current. This is illustrated in Fig.\ \ref{fig:fig3}.

The saturation is one important phenomenological difference between our
model and that of Sigrist {\it et al.}.

We want an estimate of the average twin width at which s wave surface
coherence begins in terms of experimentally measurable quantities. Roughly
speaking, this threshold will occur when the strength of the s-wave coupling
between twins is equal to the gradient energy involved in rotating the
s-wave phase by $\pi/2$ at the surface. We will assume an very simple model
with a surface layer of depth s in which $|\Psi_d|= |\Psi_d^0|$ for $z < -s$
and $|\Psi_d|=0$ for $-s<z<0$. $|\Psi_s|$ is assumed constant for all z.

The first task is to get some idea of the strength of s-wave coupling across
the twin boundary. If we consider the situation far from the surface we may
take $|\Psi_d(z=-\infty)|$ to be large and fixed. An effective free energy
may then be written down for $\Psi_s$ and an Euler-Lagrange equation for the
variation in $\Psi_s$ with respect to x derived. 
\begin{equation}
\alpha_s \Psi_s + \alpha_{sd}\Psi_d - K_s \frac{\partial^2 \Psi_s} {\partial
x^2} = 0
\end{equation}
If we assume a step function boundary where $\alpha_{sd}(x)= -sgn(x)
\alpha_{sd}^0 $ then we have the solution 
\begin{equation}
\Psi_s(x) = sgn(x)\Psi_s^0 (1-e^{-|x|\xi_s})
\end{equation}
where $\Psi_s^0$ is the bulk value. The result for the free energy per unit
area of the twin boundary is: 
\begin{equation}
\frac{F_{b}}{A} = \xi_s \alpha_s |\Psi_s^0|^2
\end{equation}

The c-axis gradient energy is also required and is roughly 
\begin{equation}
\frac{F_{g}}{A} = K_s |\Psi_s|^2 (\frac{\pi/2}{s})^2 w
\end{equation}
where $w$ is the twin width. Noting that $K_s/ \alpha_s = \xi_c^2$ and
setting $F_g=F_b$ we obtain: 
\begin{equation}
w = \frac{\xi_{ab}}{\xi_c^2} s^2
\end{equation}

For a surface layer with a depth of $100 \AA$ (about 8 unit cells) then we
obtain a twin width of approximately $1 \mu m$. We emphasize that this is
merely an order of magnitude estimate. In addition, it is not clear exactly
how deep such a surface layer should be. However, the resulting twin width
is not unreasonable. A highly twinned sample of $0.5 mm$ may have as more
than $10^3$ twins resulting in an average twin width of a few tenths of
microns. Thus while we expect no net Josephson current in a lightly twinned
sample, our model predicts the net Josphson current observed in more heavily
twinned crystals.

\section{Proposed Experiment}

Our model predicts a nonzero Josephson current resulting from a surface
effect. For samples with relatively large twins, we expect a d+s d-s
alternation between twins at the surface and no net Josephson current. This
explains why experiments on two twin crystals show a sign change in the
Josephson coupling to Pb across the twin boundary \cite{kouz}. In a sample
with many smaller twins, however, the coupling between twins wins out and a
coherent s-wave surface layer results. We expect this to take place in
samples where the average twin width is less than a few micrometers. The
s-wave surface layer is $\pi/2$ out of phase with the bulk d-wave phase.

\begin{figure}
\begin{center}
\leavevmode
\epsfxsize \columnwidth
\epsffile{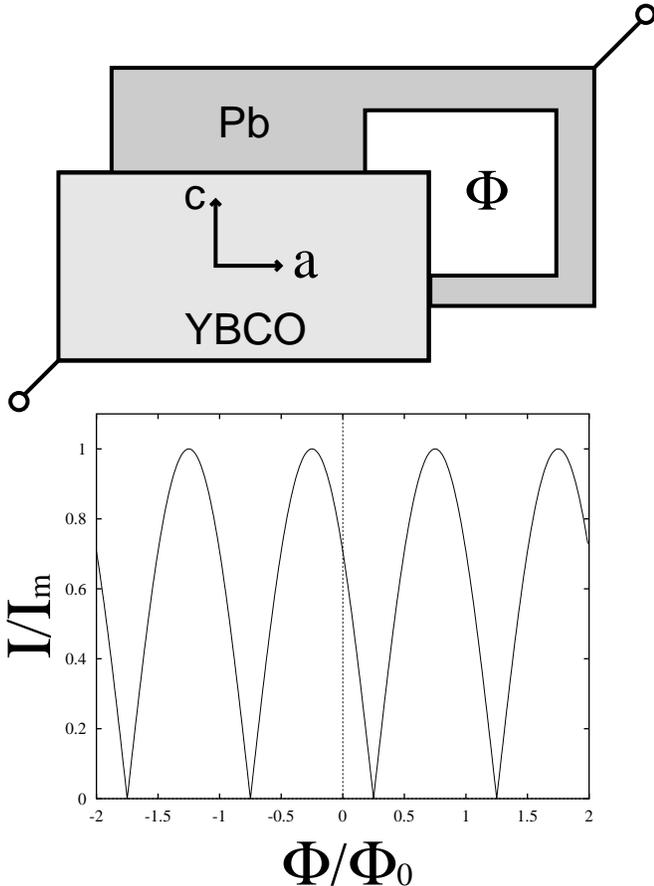}
\caption[]{Schematic of proposed YBCO-Pb SQUID experiment.  One
junction is on the (100) surface of the YBCO, the other on the
(001) surface.  The $\pi/2$ shift between the bulk d-wave
phase and the (001) surface s-wave phase shifts the current maximum
by a quarter of a flux quantum.}
\label{fig:fig4}
\end{center}
\end{figure}

We emphasize this fact because the $\pi /2$ phase shift is experimentally
verifiable. A YBCO-Pb corner junction type experiment with one junction on
the (100) surface and the other on the (001) surface of a {\it highly twinned%
} YBCO sample should be able to detect this $\pi /2$ phase shift, as was
previously suggested by Sigrist {\it et al.} \cite{sigrist}. We give a
schematic diagram of the proposed experimental configuration in Fig.\ \ref
{fig:fig4}. The current maximum as a function of field will be shifted by a
quarter of a flux quantum. The Josephson coupling to the Pb at the (100)
junction is predominantly due to the YBCO d-wave component since d-wave
suppression is not expected at this surface. Since the c-axis Josephson
coupling results from the smaller s-wave component, it is much weaker than
the a-axis coupling. The (001) junction should therefore have a much larger
area than the (100) junction in order to minimize any DC offset of the
interference pattern.

\section{Conclusion}

The present theory can reconcile the puzzles mentioned at the outset. \ The
suprprisingly large value of $I_{c}R_{n}$ is ascribed to the partial,
coherence of the s-wave component at the surface. \ The fact this coherence
is only partial gives a reasonable account of the overall differences
between single crystal and twinned samples. \ The fact that single-boundary
junctions always show a change in sign of the s component is also consistent:
in this case the twins are larger. \ These experiments also show excess
current at zero applied magnetic field. \ This would be consistent with some 
{\it partial} coherence of the s component across the boundary, as the
larger (stronger) of the two twins appears to control the weaker one. \
Hence we believe that the theory can account for all observations. \ The
experiment of the previous section would be a critical test of the theory. \
Experiments in which the relative twin populations are precisely
controlled would serve to rule out the alternative explanation in which the
current is due to accidental anisotropy introduced in the growth process. \
\ 

One important qualitative conclusion about the underlying physics of the
bulk can be drawn from this picture: s-wave competes with d-wave in YBCO. \
If our model is correct, then the naive estimate of 10\% admixture of s-wave
as a proportion of d-wave remains roughly correct. \ Expressed in the
language of Eq. \ref{fe}, we have that $|\Psi _{s}|/|\Psi _{d}|\backsim
\alpha_{sd}/(\alpha_{s}-\alpha_{d})\backsim 0.1$ at low temperatures.  \ If s-wave were
very strongly suppressed by a large positive $\alpha_{s}$, it would not be so
easily induced by the lattice distortion. \ 

The present theory would predict that only those materials with orthorhombic
distortion should show c-axis tunneling. \ Recently, c-axis Josephson
tunnneling between Ba$_{2}$Sr$_{2}$CaCu$_{2}$O$_{8+x}$ and $Pb$ \cite{Mossle}
has been observed in spite of the absence of an orthorhombic distortion in
this material. \ However, due to the fact that $I_{c}R_{n}\sim 1\mu eV$,
orders of magnitude less than the gap value, we believe that this
interesting effect is physically different from that seen in the YBCO
experiment.

We would like to thank J. Betouras for many helpful discussions. This work
was supported by the NSF under the Materials Theory Program (DMR-9704972)
and under the Materials Research Science and Engineering Center Program,
(DMR-96-32527).

\end{document}